\documentclass[aps,pra,amsfonts,floatfix,superscriptaddress,twocolumn,longbibliography,nofootinbib,10pt]{revtex4-1}

\usepackage{graphicx}
\usepackage{dcolumn}
\usepackage{bm}
\usepackage{amsmath}
\usepackage{amssymb}
\usepackage{bbold}
\usepackage{physics}
\usepackage{mathtools}
\usepackage{multirow}
\usepackage{hhline}
\usepackage{siunitx}
\usepackage{amsthm}
\usepackage{adjustbox}
\usepackage{soul,color}
\usepackage{xcolor}
\usepackage[hidelinks, colorlinks=true, allcolors=blue]{hyperref}

\newcommand{\changed}[1]{\textcolor{black}{#1}}

\usepackage{booktabs}

\usepackage{subcaption}

\usepackage[T1]{fontenc}
\renewcommand{\selectlanguage}[1]{}

\begin{document}
\preprint{APS/123-QED}
\title{Quantum Simulation via Stochastic Combination of Unitaries}

\author{Joseph Peetz}
 \email{peetz@ucla.edu}
\affiliation{Department of Physics and Astronomy, University of California, Los Angeles, CA, USA}

\author{Scott E. Smart}
\affiliation{College of Letters and Science, University of California, Los Angeles, CA, USA}

\author{Prineha Narang}
\email{prineha@ucla.edu}
\affiliation{College of Letters and Science, University of California, Los Angeles, CA, USA}

\date{\today}

\begin{abstract}
Quantum simulation algorithms often require numerous ancilla qubits and deep circuits, prohibitive for near-term hardware. We introduce a framework for simulating quantum channels using ensembles of low-depth circuits in place of many-qubit dilations. This naturally enables simulations of open systems, which we demonstrate by preparing damped many-qubit GHZ states on ibm\_hanoi. The technique further inspires two Hamiltonian simulation algorithms with \changed{gate counts that are asymptotically independent} of the spectral precision \changed{target}, reducing resource requirements by several orders of magnitude for a benchmark system.
\end{abstract}

\maketitle

\section{Introduction}
Linear combination of unitaries (LCU) has proven to be an essential primitive in the design of quantum algorithms \cite{childs_hamiltonian_2012, childs_quantum_2017, low_hamiltonian_2019, chakraborty_power_2019, wan_block-encoding-based_2021, wang_fast_2021, nguyen_block-encoding_2022, an_linear_2023}, enabling the simulation of general operators within a larger, dilated Hilbert space \cite{stinespring_positive_1955, paulsen_completely_2003, hu_quantum_2020, langer_b_1972}. While powerful as a theoretical tool, this ancilla-based dilation is often expensive in practice, generating extensive sequences of unitary operators controlled on many qubits. Combined with its need for post-selective measurements \cite{childs_hamiltonian_2012}, in total, LCU can require many repetitions of deep quantum circuits, infeasible for \changed{today's} quantum computers. 

In this work, we harness a simple stochastic framework for mapping general quantum channels onto single- or no-ancilla circuits, avoiding the expensive many-qubit dilations and post-selection costs of LCU. The stochastic combination of unitaries (SCU) method decomposes a channel into a convex combination of simple transformations, which can then be efficiently sampled and run as independent circuits. This results in low-depth circuits at the cost of additional measurement overhead, \changed{a valuable trade-off} for NISQ-era simulations. \changed{The notion of applying stochastic procedures to modify traditional LCU has recently been applied to multi-product formulas \cite{faehrmann_randomizing_2022} and general functions of Hermitian operators $f(H)$ \cite{chakraborty_implementing_2024} for applications in Hamiltonian simulation. Building on these ideas, this work defines a broadly applicable alternative to LCU, enabling the simulation of non-Hermitian Kraus operators as well as arbitrary, non-physical transformations.} 

\changed{This technique naturally applies to the simulation of open quantum systems, in which interactions with an environment induce non-unitary evolution \cite{breuer_theory_2007, head-marsden_quantum_2021, schlimgen_quantum_2021, kamakari_digital_2022}. Such processes are physically ubiquitous and have a variety of applications across quantum chemistry, condensed matter, noise modeling, and beyond \cite{egorova_modeling_2001, okolowicz_dynamics_2003, rotter_review_2015, dellesite_molecular_2017, moueddene_realistic_2020}. Simulating open systems is, in general, classically hard, and a number of quantum computational approaches have been proposed. Most techniques rely on dilating the system space to a larger Hilbert space, incurring the aforementioned resource overheads \cite{stinespring_positive_1955, langer_b_1972, paulsen_completely_2003, childs_hamiltonian_2012}. In comparison, the low-depth, qubit-efficient circuits of SCU offer significant practical advantages. To demonstrate this method, we simulate a noisy, eight-qubit entanglement generation process with high accuracy on ibm\_hanoi.}

\changed{Another critical application of SCU is Hamiltonian simulation. This is one of the most promising applications of quantum computation,} not only for quantum dynamics \cite{wiebe_simulating_2011, miessen_quantum_2023, ollitrault_molecular_2021} but also as a subroutine in ground state estimation \cite{aspuru-guzik_simulated_2005, dong_ground-state_2022, wang_quantum_2023, nam_ground-state_2020}, material characterization \cite{bauer_quantum_2020, lordi_advances_2021, de_leon_materials_2021}, and combinatorial optimization \cite{farhi_quantum_2014, albash_adiabatic_2018}. Despite substantial research into Suzuki-Trotter product formulas \cite{lloyd_universal_1996, suzuki_general_1991, childs_theory_2021, childs_nearly_2019}, post-Trotter methods \cite{low_optimal_2017, low_hamiltonian_2019}, and randomized algorithms \cite{campbell_random_2019, faehrmann_randomizing_2022}, Hamiltonian simulation remains \changed{largely} out of reach for \changed{today's} quantum computers. 

\changed{Towards reducing this gap, we introduce two simple strategies for Hamiltonian simulation with stochastic combinations of unitaries. The first approach has a close relation with the single-ancilla approach of Ref. \cite{chakraborty_implementing_2024} but realizes a simpler distribution via a unitary decomposition with tighter bounds. The second approach highlights how stochastic approaches can be incorporated into traditional procedures, reducing the error of deterministic product formulas with a stochastically treated residual. In comparison to the well-known qDRIFT algorithm \cite{campbell_random_2019}, we sample distinct, higher-order expansions that lead to improved resource scaling. Importantly, our algorithms' circuit depths are asymptotically independent of the target precision, which can enable ultra-precise simulations. We provide resource estimates for simulating the transverse field Ising model, reducing gate counts by up to several orders of magnitude compared to existing methods.}

\section{Results}
\subsection{Theoretical Framework}
Quantum channels are completely positive, trace-preserving maps and can be represented using an operator-sum representation $\mathcal{E}(\rho)=\sum_{i} K_i \rho K_i^{\dagger}$,
where the Kraus operators $K_i$ satisfy $\sum_i K_i^{\dagger} K_i = \mathbb{1}$ \cite{nielsen_quantum_2010, audretsch_entangled_2007}. We first decompose the Kraus operators into a unitary basis $\{U_j\}$ with coefficients $\{c_j\}$. By appropriately normalizing the coefficients, we can then express the channel as a convex combination times a normalization factor $\lambda := \sum_{jk} |c_j c_k^*|$:
\begin{equation} \label{eq: channel_unitary_decomp}
     \begin{split}\mathcal{E}(\rho)&=\lambda\Big(\sum_j p_{jj} U_j^{} \rho U_j^{\dagger} \\ & +\sum_{k<j} \frac{p_{j k}}{2} (e^{i \alpha_{j k}} U_j^{} \rho U_k^{\dagger}+\text {h.c.}) \Big),\end{split}
\end{equation}
where $p_{jk} := \lambda^{-1} |c_j c_k^*| $. The channel consists of random-unitary terms ($j=k$) and coherence-preserving cross terms ($j \neq k$).

\begin{figure}[t!]
\centering
\includegraphics[width=\columnwidth]{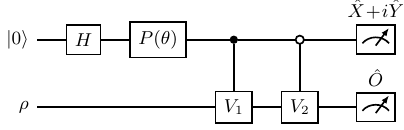} 
\caption{Quantum circuit simulating $\operatorname{Tr}\left[\hat{O} e^{i \theta} V_1 \rho V_2^\dagger \right]$, where $P(\theta)$ is a phase gate.}
\label{figure:cross-term-circuit}
\end{figure}

To simulate the channel, we can sample the terms $\left\{U_j^{} \rho U_j^{\dagger}\right\}$ and $\left\{\frac{1}{2} (e^{i \alpha_{j k}} U_j^{} \rho U_k^{\dagger}+ \text{h.c.})\right\}$ with their respective probabilities. The random-unitary terms map directly onto circuits consisting of $U_j$ acting on the input state \cite{peetz_simulation_2024}, and the cross terms map onto circuits of the form shown in Fig. \ref{figure:cross-term-circuit}. In Eq. \eqref{eq: channel_unitary_decomp}, each cross term and its Hermitian conjugate can be accessed simultaneously via a single $\hat{X}$ measurement of the ancilla qubit \cite{faehrmann_randomizing_2022}.

This stochastic combination of unitaries (SCU) thus generates an ensemble of circuits, with considerably simpler structure and depth compared to linear combination of unitaries (LCU) \cite{childs_hamiltonian_2012}. Repeatedly sampling this ensemble for $N$ samples yields an unbiased estimator of the quantum channel \cite{peetz_simulation_2024, arrasmith_operator_2020}, with variance scaling as $\mathcal{O}(\lambda^2/N)$. In the special case of random-unitary channels, Eq. \eqref{eq: channel_unitary_decomp} has unit norm ($\lambda=1$) and excellent scalability \cite{peetz_simulation_2024}. For general channels, we can mitigate the increased variance via additional measurements, scaling up the number of samples $N$ by $\lambda^2$. For depth-limited devices, this trade-off between low-depth circuits and additional measurement overhead can enable otherwise infeasible simulations.

\changed{To precisely bound the number of shots required, we adapt error analysis by Faehrmann et al. for randomizing multi-product formulas \cite{faehrmann_randomizing_2022}. Given an observable $O$ and a target state $\rho$, we seek to estimate $\Tr[O \rho]$ with statistical precision $\epsilon_{stat}$. Applying Hoeffding's inequality, we need the following number of shots \cite[Eq. ~(24)]{faehrmann_randomizing_2022}:
\begin{equation} \label{eq: SCU-number-samples}
N_{\mathrm{SCU}} \;\ge\; \frac{2\,\lambda^{2}\,\|O\|^{2}}{\epsilon_{stat}^{2}}\,
\ln\!\frac{2}{\delta},
\end{equation}
where $\|O\|$ denotes the spectral norm of the observable and $\delta\in(0,1)$ is the probability of failure. This expression determines the sampling cost of SCU-based simulation algorithms, where $\lambda$ is the normalization constant in Eq. \eqref{eq: channel_unitary_decomp}.}

\changed{In comparison, by coherently encoding the unitary coefficients as amplitudes of ancilla registers, linear combination of unitaries achieves a quadratic scaling reduction in $\lambda$. Specifically, Cleve et al. introduce an LCU-based method for simulating quantum channels in which the probability of successful post-selection scales as $1/\lambda$ \cite[Eq. ~(19)]{cleve_lindblad_2019}. Combined with the error analysis of Faehrmann et al. \cite{faehrmann_randomizing_2022}, this approach requires the following number of shots:
\begin{equation}
N_{\mathrm{LCU}} \;\ge\; \frac{2\,\lambda\,\|O\|^{2}}{\epsilon_{stat}^{2}}\,
\ln\!\frac{2}{\delta}.
\end{equation}
However, this scaling improvement comes at the cost of a potentially significant increase in circuit depth. Suppose that a channel's $m$ Kraus operators $K_i$ each decompose into $M_i$ unitaries, and let $q = \max_i  \{M_i\}$. Then, the ``select'' stage of LCU requires $\sum_i M_i$ controlled operators and a total of $\lceil\log_2 m \rceil + \lceil\log_2 q \rceil$ ancilla qubits \cite{cleve_lindblad_2019}. In contrast, the cost of SCU has no direct dependence on the number of unitaries $M_i$. Note that the dependence of $N_{\mathrm{LCU}}$ on $\lambda$ can be eliminated entirely through oblivious amplitude amplification (OAA). However, this further increases the circuit depth by a factor of $\mathcal{O}(\sqrt{\lambda})$ \cite{brassard_quantum_2000, berry_exponential_2014, berry_simulating_2015, cleve_lindblad_2019}.}

\begin{table}[t!]
\centering
\renewcommand{\arraystretch}{1.25} 
\begin{tabular}{@{}l@{\hspace{4em}}c@{\hspace{4em}}c@{}}
\hline \hline
 Method & Depth & Shots \\
\hline
SCU & $\mathcal{O}(1)$ & $\mathcal{O}(\lambda^2)$ \\
LCU & $\widetilde{\mathcal{O}}(M)$ & $\mathcal{O}(\lambda)$ \\
LCU + OAA & $\widetilde{\mathcal{O}}(M \sqrt{\lambda})$ & $\mathcal{O}(1)$ \\
\hline\hline
\end{tabular}%
\caption{\changed{Resource comparison for simulating a single operator $A = \sqrt{\lambda} \sum_{j=1}^M p_j U_j$, where $p_j > 0$ and $\sum_j p_j = 1$. Here, we assume that each unitary $U_j$ can be implemented in $\mathcal{O}(1)$ circuit depth. Stochastic combination of unitaries (SCU) uses the techniques introduced in this paper. Linear combination of unitaries (LCU) uses the block-encoding method introduced in Ref. \cite{childs_hamiltonian_2012}. LCU with oblivious amplitude amplification (OAA) uses the techniques from Refs. \cite{brassard_quantum_2000, childs_hamiltonian_2012, berry_exponential_2014, berry_simulating_2015, cleve_lindblad_2019}.}}
\label{tab:depth-shots-table}
\end{table}

\changed{As a simple but useful reference, we compare resources for simulating a single operator acting on a state in Table \ref{tab:depth-shots-table}. We see that each method offers a different trade-off between circuit depth and number of shots, with SCU providing the lowest depth circuits but also the greatest number of shots. Collectively, these methods offer useful flexibility for different architecture implementations. Notably, in a fully fault-tolerant setting, one may prioritize total runtime over either of these metrics. Assuming each shot is run in series, we see that SCU has a lower runtime than LCU when $\lambda < \widetilde{\mathcal{O}}(M)$ and a lower runtime than LCU with OAA when $\lambda < \widetilde{\mathcal{O}}(M^{2/3})$. Qualitatively, SCU offers the greatest runtime advantage when the target operator decomposes into many low-weight unitaries. It also enables otherwise impossible simulations on qubit- and depth-limited devices.}

\subsection{Damped GHZ Simulation}
SCU naturally applies to the simulation of open quantum systems, which we demonstrate by simulating an eight-qubit GHZ state subject to CNOT-induced amplitude damping on ibm\_hanoi. The GHZ state is a common example of multipartite entanglement, often analyzed in the context of quantum networks \cite{avis_analysis_2023, meignant_distributing_2019, patil_entanglement_2022}. Realistic networks are subject to noise, such as photon loss and decoherence \cite{covey_quantum_2023}, and our simulation serves as a guide that could be readily adapted to hardware. 

\begin{figure}[t!]
\centering
\includegraphics[width=\columnwidth]{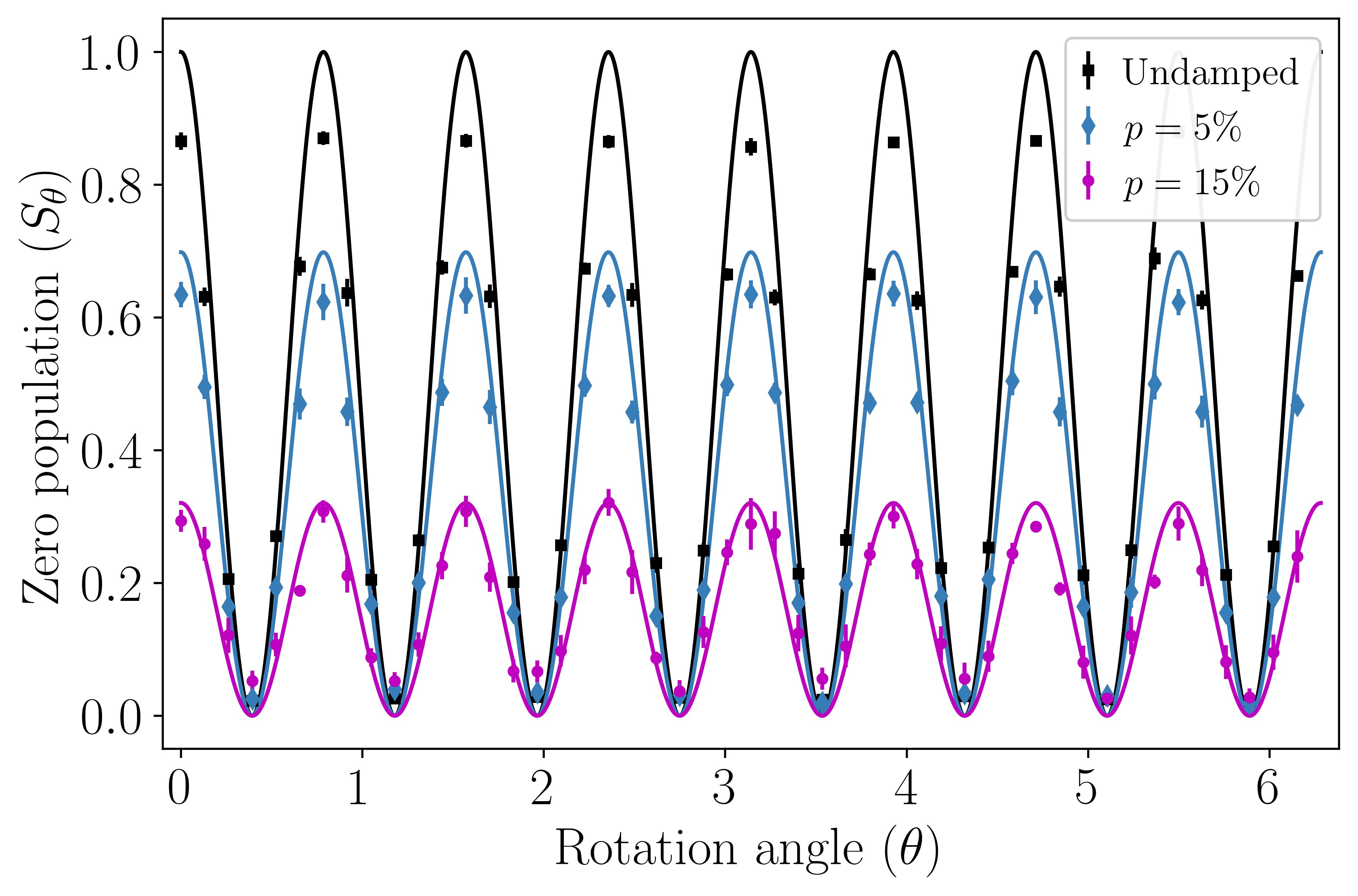} 
\caption{Eight-qubit damped GHZ simulation, plotting the all-zero MQC population $\rho_{0^n, 0^n}$ as a function of the rotation angle $\theta$ and damping strength $p$. The points are average measurements over five runs on ibm\_hanoi, each using $1000$ shots per angle, and the curves are analytically predicted signals.}
\label{fig:MQC-signal-plot}
\end{figure}

As described in Section \ref{sec:damped-GHZ-methods} of the Methods, we measure the fidelity of the damped GHZ via a modified version of multiple quantum coherences (MQC) \cite{garttner_relating_2018}. In Fig. \ref{fig:MQC-signal-plot}, we plot the MQC signal for different damping strengths $p$, comparing experimental results from ibm\_hanoi with analytically predicted signals. This simulation gives excellent theoretical agreement in spite of the additional variance induced by SCU.

We then Fourier transform these signals to compute the fidelity $F\left(\rho, \rho^{\text{GHZ}}\right)$. For damping strengths $p \in \{0,0.05,0.15\}$, we respectively measure the fidelities $\{0.932 \pm 0.007,0.76 \pm 0.01,0.539 \pm 0.008\}$ over five runs. These closely match the ideal predictions of $\{1,0.842,0.613\}$, up to a slight offset due to device noise. \changed{These experiments were run on ibm\_hanoi in February 2024, using an 8-qubit subset chosen by the Qiskit transpiler.}

Compared to a direct LCU approach, SCU results in much simpler circuits, with reasonable sampling overhead for small damping parameters. Furthermore, the single-qubit amplitude damping channel has known one-qubit dilations that outperform direct LCU. For example, the approach in \cite{rost_simulation_2020} compiles to four CNOT gates for each instance of damping, compared to $\frac{2p}{1+p}$ CNOT gates on average with SCU. Concretely, in the eight-qubit GHZ simulation, the base MQC circuit with no damping requires 14 CNOTs. To include damping at $p=15\%$, the implementation in \cite{rost_simulation_2020} would require $56$ additional CNOTs compared to only $3.7$ on average with SCU, a massive improvement that enables an otherwise prohibitive simulation. We provide further resource analysis in \changed{\ref{appendix-damping-resources}}.

\subsection{Hamiltonian Simulation via Convex Taylor Sampling}
Building on this stochastic framework, we introduce two algorithms for Hamiltonian simulation: (1) convex Taylor sampling (CTS) and (2) stochastically enhanced product formulas. In each algorithm, we expand time evolution as a probabilistic combination of quantum operations,
\begin{equation} \label{eq: hamsim-left-right-convex}
\begin{aligned}
    e^{-iHt} \rho(0) e^{iHt} = \lambda \Biggl(\sum_j p_j U_j\Biggr) \rho(0) \Biggl(\sum_k p_k U_k^\dagger\Biggr).
\end{aligned}
\end{equation}
From this form, we approximate the channel via randomized sampling of $\{U_j\}$ and $\{U_k^\dagger\}$, obtaining left/right gate pairs for each time step. These transformations are then simulated via SCU with measurement overhead $\lambda^2$. While many such expansions exist \cite{faehrmann_randomizing_2022, chakraborty_implementing_2024}, we propose two decompositions of the time evolution operator which simultaneously optimize the spectral precision $\epsilon$ and the normalization constant $\lambda$. Remarkably, both of our algorithms have gate complexities that are asymptotically independent of the target precision, thus enabling simulations with low spectral error requirements.

The following theorem provides a simple decomposition of the Taylor series of a time evolution operator, with inspiration from Ref.~\cite{wan_randomized_2022}. We use this decomposition to construct a simulation algorithm, with step-by-step instructions in Fig. \ref{fig:CTS-pseudocode}. \\

\begin{figure}[!t]
    \centering
    \adjustbox{valign=t, margin=1em, minipage=[c]{0.9\columnwidth}, frame, center}{\raggedright \textbf{Input:} A time-independent Hamiltonian decomposed \\ \quad \quad \quad \; into Pauli strings, $H = \sum_i c_i P_i$. \\
    \vspace{1.5mm}
    \textbf{Output:} A sequence of left/right gate pairs.
    \begin{enumerate}
    \item Discretize the time evolution into $r$ time steps,
    $$ \rho(t) = \left(e^{-iHt/r}\right)^r \!\rho(0) \left(e^{iHt/r}\right)^r. $$
    \item Classically compute a convex approximation of $e^{-iHt/r}$ to truncation order $M$:
    $$\mathcal{C}_{\text{T}}^{(M)}(t/r) = \sum_j p_j U_j.$$ \vspace{-3mm}
    \item Repeatedly sample $\mathcal{C}_{\text{T}}^{(M)}(t/r)$ and $\mathcal{C}_{\text{T}}^{(M)}(-t/r)$ to generate $r$ left/right pairs of quantum gates.
\end{enumerate}}
    \caption{Hamiltonian simulation via stochastic combination of unitaries.}
    \label{fig:CTS-pseudocode}
\end{figure}

    
    

\noindent \textbf{Theorem 1}: For a Hermitian operator $H = \sum_i c_i P_i$ and real parameter $t$,
\begin{equation}
    e^{-iHt} = L_c \sum_j p_j P'_j + \sqrt{1+L_s^2} \sum_k p_k e^{i \theta P'_k},
\end{equation}
where $\sum_i p_i = 1$, $\theta = \arcsec{(\sqrt{1+L_s^2})}$, and $P'_i \in \pm \{I,X,Y,Z\}^{\otimes n}$ represent Pauli strings up to a negative sign. Further, $L_c$ and $L_s$ are bounded as $L_c \leq \cosh{(t |H|_1)} - 1$ and $L_s \leq \sinh{(t |H|_1)}$, where $|H|_1 := \sum_i |c_i|$.

\noindent \textbf{Proof}: See \ref{subsection: appendix-theorem1-proof}. \\



Theorem 1 expresses the Taylor series as a convex combination of unitary operators up to an overall $L_1$ norm. This form lends itself to convex Taylor sampling (CTS), an algorithm that leverages SCU to independently sample the left and right propagators, $e^{-iHt}$ and $e^{iHt}$. Specifically, for a time step $t$, we truncate the expansion in Theorem 1 to order $M$ and normalize by the overall $L_1$ norm $\mu(t) = L_c+\sqrt{1+L_s^2}$. We then sample the resulting convex combination
\begin{equation} \label{eq: CTS-convex-combination}
\mathcal{C}_{\text{T}}^{(M)}(t) = \sum_j p'_j P'_j + \sum_k p'_k e^{i \theta P'_k},
\end{equation}
where the coefficients $p'_j := \frac{L_c}{\mu(t)} p_j$ and $p'_k := \frac{\sqrt{1+L_s^2}}{\mu(t)} p_k$ form a probability distribution. Independently, we sample $\mathcal{C}_{\text{T}}^{(M)}\!\left(-t\right)$ to obtain an overall transformation of the form $V_1 \rho V_2^\dagger$, as implemented in Fig. \ref{figure:cross-term-circuit}. To minimize simulation error, we discretize into $r$ time steps as follows:
\begin{equation} \label{eq: CTS-spectral-error}
e^{-iHt} = \sqrt{\lambda} \left(\mathcal{C}_{\text{T}}^{(M)}\!\left(\frac{t}{r}\right)\right)^r + \mathcal{O}\left(\frac{(|H|_1 t)^{M+1}}{r^M}\right),
\end{equation}
where $\lambda = \mu(t/r)^{2r}$ is the normalization constant for the full channel. We see that for a tolerated error $\epsilon$, CTS requires $r \sim \mathcal{O}((|H|_1 t)^{1+1/M} \epsilon^{-1/M})$ time steps.

\begin{figure*}[t!]
    \centering
    \begin{subfigure}[b]{0.3394\textwidth}
        \includegraphics[width=\textwidth]{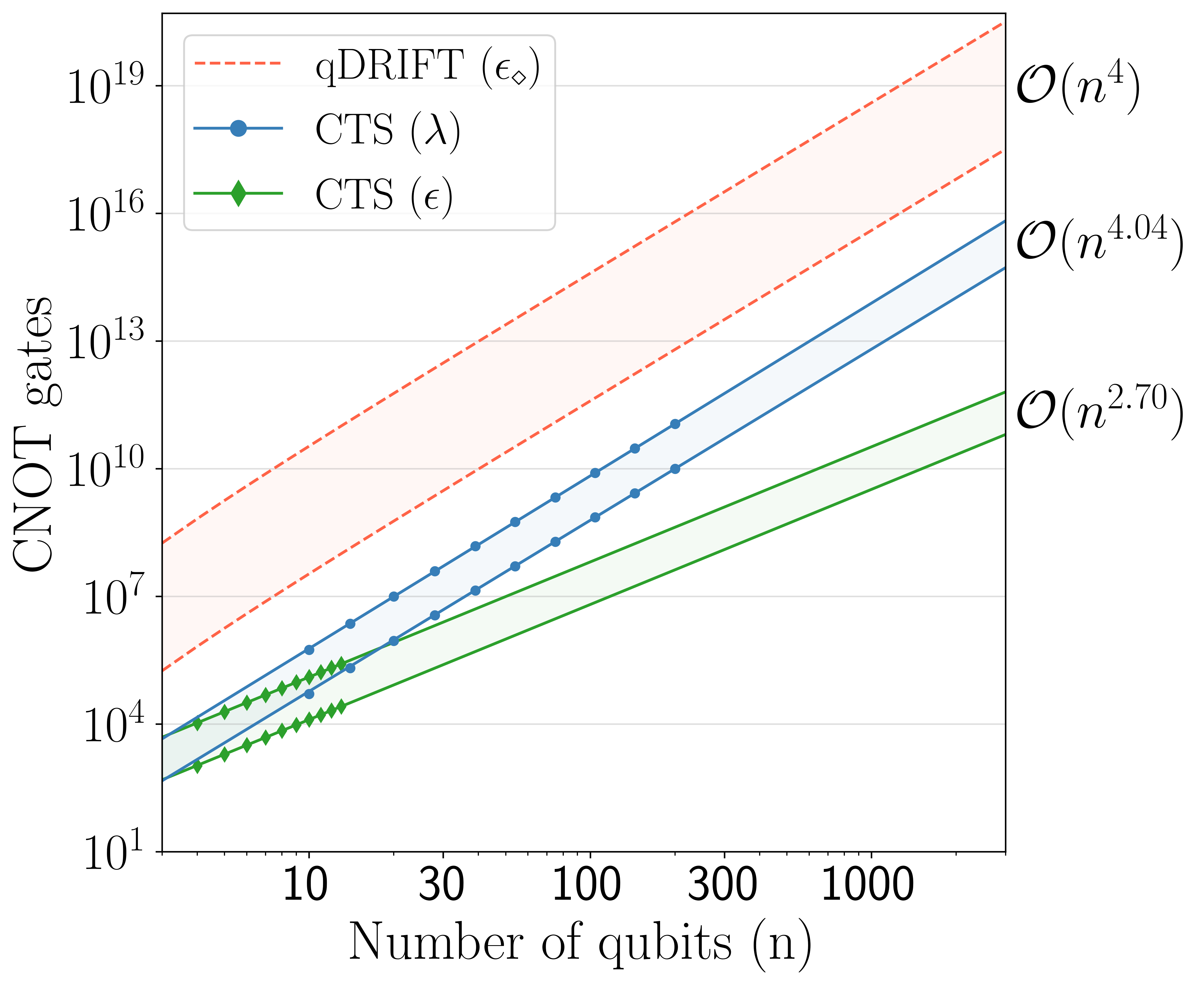}
        \caption{}
        \label{subfig:TFIM_CNOT_counts:qdrift}
    \end{subfigure}
    \begin{subfigure}[b]{0.3238\textwidth}
        \includegraphics[width=\textwidth]{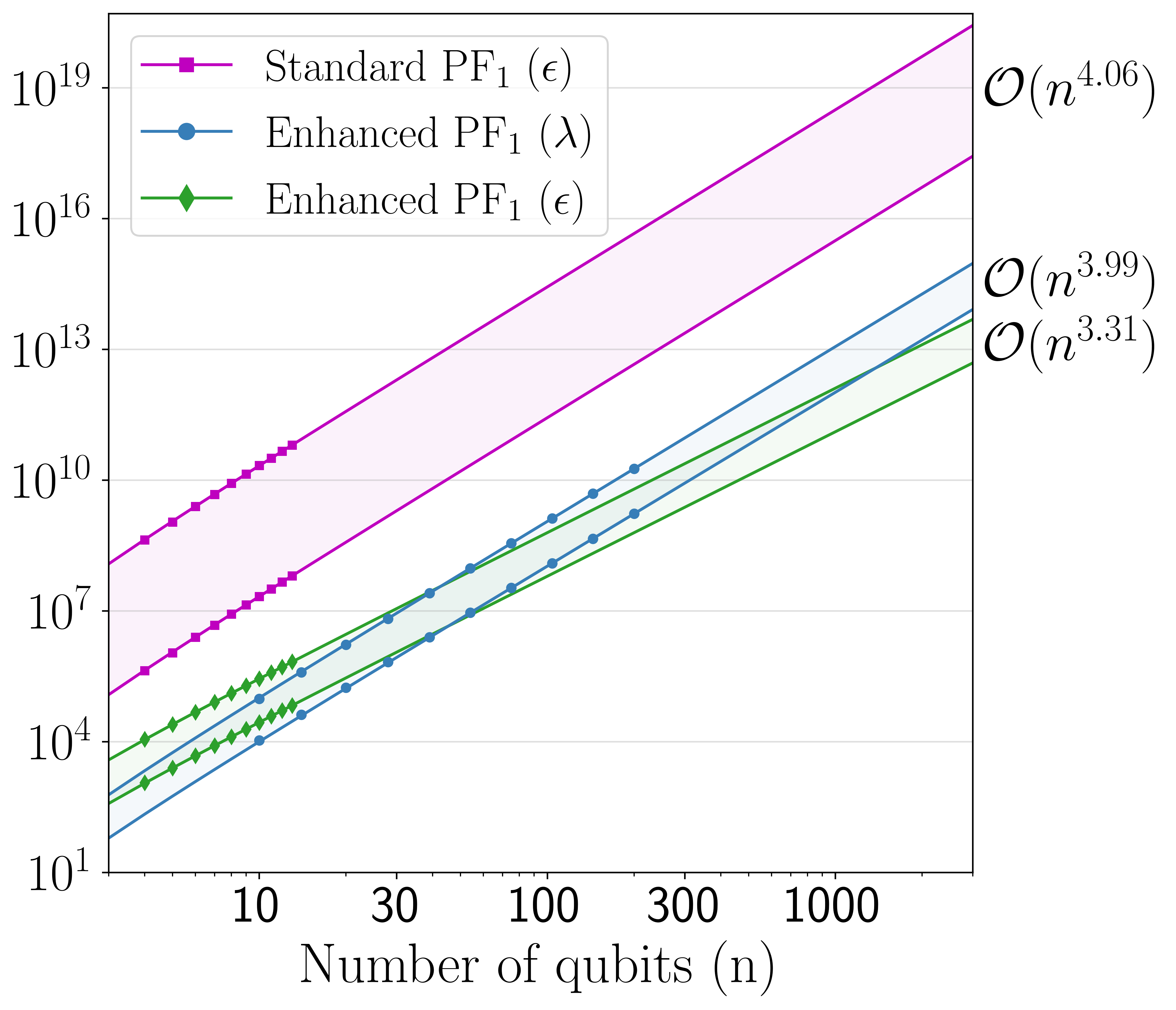}
        \caption{}
        \label{subfig:TFIM_CNOT_counts:pf1}
    \end{subfigure}
    \begin{subfigure}[b]{0.3238\textwidth}
        \includegraphics[width=\textwidth]{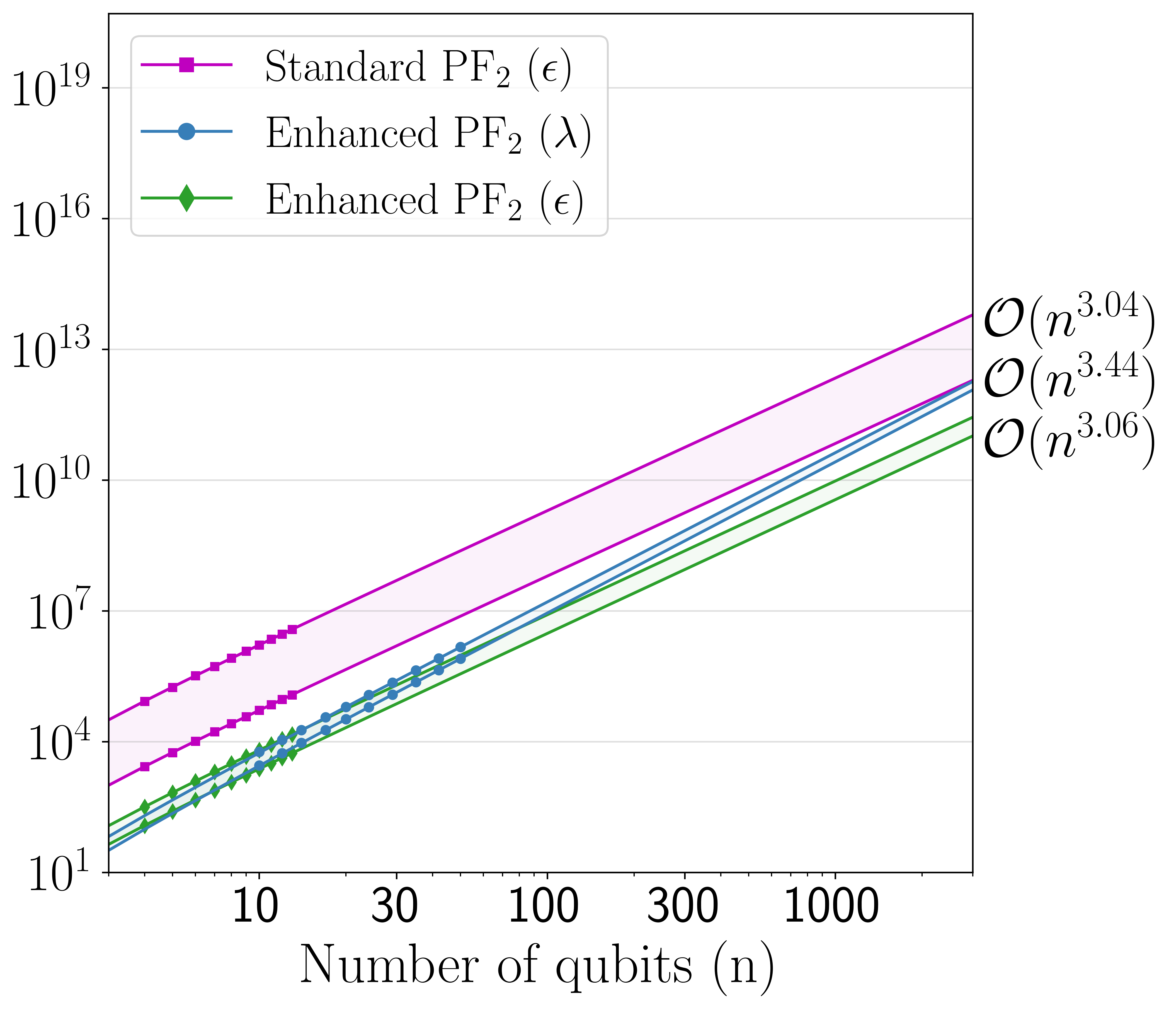}
        \caption{}
        \label{subfig:TFIM_CNOT_counts:pf2}
    \end{subfigure}
    \caption{Resource estimates for simulating the TFIM for different system sizes $n$, with simulation time $t=n$. We respectively compare (a) qDRIFT to CTS, (b) the first-order product formula to its enhanced version, and (c) the second-order product formula to its enhanced version. \changed{We used a truncation order of $M=3$ for CTS and the enhanced first-order product formula and $M=7$ for the enhanced second-order product formula.} For qDRIFT, we use analytic estimates \cite{campbell_random_2019} and the diamond norm error range $2 \times 10^{-6} \leq \epsilon_\diamond \leq  2 \times 10^{-3}$. For all other algorithms, the points are empirically calculated bounds, and the lines are power law fits. The magenta and green bands are bounded by $10^{-6} \leq \epsilon \leq  10^{-3}$, and the blue bands are bounded by $2 \leq \lambda \leq 2000$.}
    \label{fig:TFIM_CNOT_counts}
\end{figure*}

To mitigate the additional variance induced by the normalization constant $\lambda$, we budget a maximum measurement overhead, creating a second constraint on the number of time steps. Because the leading-order contribution to $\lambda$ is quadratic, we get
\begin{equation} \label{eq: CTS-lambda}
\lambda \sim 1 + \mathcal{O}\left(\frac{(|H|_1 t)^2}{r}\right).
\end{equation}
Thus, to limit the measurement overhead to $\lambda^2$, we require $r \sim \mathcal{O}((|H|_1 t)^{2}(\lambda-1)^{-1})$ time steps. For $M > 1$, this overhead constraint is asymptotically dominant over the constraint on the spectral precision.

\subsection{Hamiltonian Simulation via Stochastically Enhanced Product Formulas}
The Suzuki-Trotter product formulas are perhaps the most well-known algorithms for Hamiltonian simulation \cite{lloyd_universal_1996, suzuki_general_1991}. For the Hamiltonian $H = \sum_j c_j P_j$, the first-order product formula approximates its time evolution to $\mathcal{O}(t^2)$ as
\begin{equation}
\mathcal{S}_1(t) = \prod_{j=1}^m e^{-i c_j P_j t}.
\end{equation}
More generally, $\mathcal{S}_p(t)$ denotes the $p$-th order product formula. By construction, higher-order product formulas correct asymmetries from the operators $P_j$ not commuting.

We can enhance these formulas by stochastically implementing their remainders as higher-order error corrections. For example, the leading error of the first-order product formula is a sum of commutators, $\frac{t^2}{2} \sum_{i<j} \left[P_i, P_j\right]$. More generally, we can compute the remainders of a $p$-th order product formula up to a truncation order $M$. This yields a convex combination up to the normalization constant $\mu(t)$:
\begin{equation} \label{eq: enhanced-PF-convex-combination}
\mathcal{C}_{\mathcal{S}_p}^{(M)}(t)  = p_0 S_p(t) + \sum_k p_k P_k',
\end{equation}
where the Pauli strings are computed from the remainder expressions. We discuss a procedure for computing the remainder in Section \ref{sec:remainder} of the Methods. Similarly to CTS, we can then leverage SCU to randomly simulate this expansion, as described in Fig. \ref{fig:CTS-pseudocode}. For a desired precision $\epsilon$, this stochastically enhanced product formula requires $r \sim \mathcal{O}(t^{1+1/M} \epsilon^{-1/M})$ time steps.

Once again, the overall normalization constant $\lambda = \mu(t/r)^{2r}$ determines our sampling overhead, with asymptotic scaling determined by the leading-order remainder. For the enhanced $p$-th order product formula,
\begin{equation}
\lambda_{\mathcal{S}_{p}} \sim 1 + \mathcal{O}\left(\frac{\alpha^{(p)}_{\text{comm}}t^{p+1}}{r^{p}}\right),
\end{equation}
where $\alpha^{(p)}_{\text {comm}}$ is the $L_1$ norm of the nested commutator expression in Ref.~\cite{childs_theory_2021}. Thus, we require $r \sim \mathcal{O}((\alpha^{(p)}_{\text{comm}})^{1/p} t^{1+1/p}(\lambda-1)^{-1/p})$ time steps to bound the measurement overhead to $\lambda^2$. This is asymptotically dominant over the constraint on spectral precision.

\subsection{Transverse Field Ising Simulation}
To highlight the performance of these approaches, we compare gate costs for simulating the transverse field Ising model (TFIM). The TFIM Hamiltonian is $H = -J \sum_{i} Z^{(i)} Z^{(i+1)} - h \sum_j X^{(j)}$, where $J$ is the exchange interaction parameter and $h$ quantifies the strength of the transverse magnetic field.

We compute the number of CNOT gates required to simulate the TFIM for system size $n$, evolution time $t=n$, and model parameters $J=1$ and $h=1$, following Ref.~\cite{childs_toward_2018}. To compare dependencies on spectral precision $\epsilon$ and normalization constant $\lambda$, we consider the parameter ranges $10^{-6} \leq \epsilon \leq  10^{-3}$ and $2 \leq \lambda \leq 2000$. Our resource estimates for all algorithms are combined in Fig. \ref{fig:TFIM_CNOT_counts}. For a broad range of sizes, our algorithms reduce the CNOT counts by several orders of magnitude. This difference is particularly pronounced for simulations with low spectral error, i.e., $\epsilon \leq 10^{-6}$, due to our algorithms' asymptotic independence of spectral precision. Importantly, for many parameter choices, the reductions in depth exceed the measurement overhead of $\lambda^2$, \changed{thus} improving the overall runtime. 

\changed{In Fig. \ref{fig:TFIM_CNOT_counts}, the total CNOT costs are the product of the average gates per time step and the requisite number of time steps. While the standard and enhanced PFs differ only logarithmically in the gates per step (see \ref{appendix-TFIM-resource-analysis}), the enhanced PFs achieve substantially smaller errors per step. As a result, they require significantly fewer time steps, which is the main source of the savings observed in Figs. \ref{subfig:TFIM_CNOT_counts:pf1} and \ref{subfig:TFIM_CNOT_counts:pf2}.}

\section{Discussion}
The SCU approach provides a general framework for realizing quantum channels as ensembles of simple quantum circuits. SCU repeatedly samples a sub-normalized channel's convex unitary decomposition, yielding unitary gates for the random-unitary terms and single-ancilla dilations for the cross terms. This approach \changed{can mitigate} the need for the large ancilla networks and multi-controlled gates required in techniques like linear combination of unitaries \cite{childs_hamiltonian_2012}, quantum signal processing \cite{low_optimal_2017}, and qubitization \cite{low_hamiltonian_2019}. Instead, SCU generates low-depth circuits at the cost of additional measurements, a valuable trade-off for near-term quantum devices.

We first demonstrated SCU in the context of open quantum systems, simulating an eight-qubit, noisy GHZ state with high accuracy on ibm\_hanoi. SCU significantly reduced circuit depths while maintaining reasonable measurement costs, enabling an otherwise prohibitive simulation. Beyond this example, one could similarly model time-dependent bath interactions via sequential sampling of time-dependent unitary decompositions.

In applying SCU to the problem of Hamiltonian simulation, we showed that stochastically executing high-order series operators can significantly reduce CNOT costs, with interesting parallels to composite simulation algorithms \cite{hagan_composite_2023}. These ideas can similarly apply to higher-order product formulas as well as different systems of interest, such as fermionic and power law Hamiltonians. The algorithms' asymptotic independence of target precision is particularly attractive for simulations requiring low spectral error. \changed{For particular applications, these Hamiltonian simulation algorithms may require ancilla management strategies tailored to hardware capabilities, as multiple applications of SCU may require distinct ancilla. Using mid-circuit measurements (MCMs), one can repeatedly reset and reuse a single ancilla qubit, but generally different qubits should be utilized (see \ref{appendix-TFIM-resource-analysis}).}

The CTS algorithm is closely related to qDRIFT \cite{campbell_random_2019}. qDRIFT approximates the time propagator to first order in time as a random-unitary channel, requiring $r \sim \mathcal{O}((|H|_1 t)^2 \epsilon^{-1})$ time steps \cite{campbell_random_2019}. While both algorithms' gate complexities scale quadratically in $|H|_1$ and $t$, CTS is asymptotically independent of the spectral precision $\epsilon$ and instead scales with the maximum tolerated normalization constant $\lambda$. This difference is crucial because $\lambda$ and $\epsilon$ typically differ by several orders of magnitude. It also relates to the algorithm proposed in Ref. \cite{chakraborty_implementing_2024}, which similarly uses the single-ancilla framework but with an $M$-nested Taylor decomposition \cite{wan_randomized_2022}. In comparison, CTS simplifies both the sampling process and the resulting gates while tightening bounds on the overhead $\lambda^2$. Explicitly calculating higher terms in practice allows for cancellations to occur, further lowering the overhead.

\changed{The measurement overhead of SCU can be substantially impacted by the choice of unitary bases and the resultant $L_1$ norm of an operator's unitary decomposition. While this work focuses on the Pauli basis $\{P_j\}$ for simplicity, future work can investigate the use of alternative bases \cite{yen_measuring_2020, loaiza_reducing_2023, wu_optimization_2024}. For example, one could consider rotated Pauli frames of the form $\{V P_j V^\dagger\}$ for an optimized or heuristically chosen unitary $V$. In general, one seeks a unitary basis that minimizes the parameter $\lambda$, as this directly increases the sampling cost.}

\section{Methods}

\subsection{Damped GHZ Simulation} \label{sec:damped-GHZ-methods}

By applying a Hadamard gate and a chain of CNOTs, one prepares the $n$-qubit GHZ state,
\begin{equation}
    \ket{\text{GHZ}} = \frac{1}{\sqrt{2}}(\ket{0^n} + \ket{1^n}).
\end{equation}
Because two-qubit gates are often error-prone in practice, we model a noisy GHZ state with a CNOT-induced amplitude damping channel, with Kraus operators $K_0 = \frac{1}{2}(1+\sqrt{1-p}) I + \frac{1}{2}(1-\sqrt{1-p}) Z$ and $K_1 =  \frac{\sqrt{p}}{2} X + \frac{i \sqrt{p}}{2} Y$. That is, after each CNOT in the preparation of $\ket{\text{GHZ}}$, we sample the convex decomposition of the quantum channel as in Eq. \eqref{eq: channel_unitary_decomp}; we then apply the sampled operators to the target qubit according to SCU. In the Pauli basis, the $L_1$ norms of $K_0 \rho K_0^\dagger$ and $K_1 \rho K_1^\dagger$ are $1$ and $p$, respectively, making the normalization constant of the channel $1 + p$. For the complete $n$-qubit GHZ preparation channel, assuming that each CNOT has equal damping strength, the overall normalization constant is thus $(1+p)^{n-1}$.

We then measure the fidelity of the prepared noisy state $\rho$ relative to an ideal state $\rho^{\text{GHZ}}$ via a modified version of multiple quantum coherences (MQC) \cite{garttner_relating_2018}, detailed in \changed{\ref{appendix-damped-MQC}}. Following the approaches of \cite{wei_verifying_2020, mooney_generation_2021}, we expand the fidelity as
\begin{equation} \label{eq: fidelity-decomp}
F\!\left(\rho, \rho^{\text{GHZ}}\right)\!=\!\frac{1}{2}\!\left(\rho_{0^n, 0^n}+\rho_{1^n, 1^n}+\rho_{0^n, 1^n}+\rho_{1^n, 0^n}\right)\!.
\end{equation}
In this sum, the all-zero and all-one populations are directly observable in the standard basis. The third and fourth terms are coherences, accessible via MQC. \changed{Specifically, the MQC circuit shown in Fig. \ref{fig: MQC-circuit} gives the signal $S_\theta = \rho_{0^n, 0^n}(\theta)$, and we obtain the coherences via the Fourier transform of this signal.} As discussed in \changed{\ref{appendix-damped-MQC}}, the inversion stage of our MQC circuit exactly mirrors the preparation phase, including any sampled damping operators. We note that this technique generalizes the use of MQC to arbitrary channels, expanding beyond its current scope \cite{garttner_relating_2018}.

For a damped GHZ state, the analytically predicted signal is the ideal GHZ sinusoid together with a damping prefactor,
\begin{equation}
S_\theta = \frac{1}{2} (1-p)^{n-1} (1+\cos (n \theta)).
\end{equation}

\subsection{Convex Taylor Sampling Procedure}\label{sec:CTS_method}

\changed{We compute the convex combination in Eq. \eqref{eq: CTS-convex-combination} using a Pauli algebra representation in Python. To arrive at the operator decomposition shown in Theorem 1, we first expand the Taylor series of $e^{-iHt}$ in the Pauli basis up to truncation order $M$. The Pauli strings with real coefficients are sampled directly, and those with imaginary coefficients are grouped into Pauli rotations of the form $e^{i \theta P_k'}$. The exact process is detailed in the proof of Theorem 1 in \ref{subsection: appendix-theorem1-proof}. For Hamiltonians with $K$ terms, computing $\mathcal{C}_{\text{T}}^{(M)}(t)$ has a classical time complexity of $\mathcal{O}(K^M)$, limiting us to fairly low truncation orders $M$ for large system sizes. However, the Markov sampling techniques discussed in \ref{appendix-markov-sampling} \space enable us to use truncation orders as high as $M=7$ in our simulations.}

\subsection{Calculating Remainder Formulas}\label{sec:remainder}

\changed{To compute the product formula remainder $\sum_k p_k P_k'$ in Eq. \eqref{eq: enhanced-PF-convex-combination}, we first use symbolic computations to expand the Taylor series of both $e^{-iHt}$ and the product formula $S_p(t)$. We then expand the difference in the Pauli basis and simplify. For example, if $H = A + B$ and we use a first-order product formula $S_1(t)$, then the leading-order remainder is
\begin{equation}
    e^{-iHt} - S_1(t) = \frac{t^2}{2}(AB - BA) + \mathcal{O}(t^3),
\end{equation}
consistent with the expected commutator scaling \cite{childs_theory_2021}. To construct the sample-friendly form of Eq. \eqref{eq: enhanced-PF-convex-combination}, we then expand $A$ and $B$ in the Pauli basis and compute the right-hand side using Pauli algebra rules. Similarly to CTS, for a Hamiltonian with $K$ terms, this process incurs a classical time cost of $\mathcal{O}(K^M)$. For instance, in the example above, the dominant classical cost arises from computing the Pauli decompositions of the products $AB$ and $BA$. If $A$ and $B$ each have $\mathcal{O}(K)$ terms, the runtime of this computation is $\mathcal{O}(K^2)$.}

\subsection{Transverse Field Ising Simulation}

\changed{A practical simulation will need to simultaneously satisfy constraints on both the spectral error $\epsilon$ and the sampling overhead $\lambda^2$. Both $\epsilon$ and $\lambda$ vary with the number of time steps $r$, as shown in Eqs. \eqref{eq: CTS-spectral-error} and \eqref{eq: CTS-lambda} for CTS. In Fig. \ref{fig:TFIM_CNOT_counts}, the $\epsilon$-labeled curves are generated by finding the minimum $r$ satisfying a given spectral error budget, e.g., $\epsilon \leq 10^{-3}$. The $\lambda$-labeled curves are similarly generated by finding the minimum r satisfying a given normalization budget, e.g., $\lambda \leq 2$. In practice, we have to satisfy both constraints, choosing the larger value of $r$ from the two constraints.}

To bound the sampling overhead, we evaluate each algorithm's series expansion to truncation order $M$, using Pauli string algebra rather than a matrix representation. For CTS and the stochastically enhanced first-order product formula, we compute the respective expansions to order $M=3$ up to 200 qubits. For the enhanced second-order product formula, we compute corrections to order $M=4$ and use Markov sampling for orders $5 \leq M \leq 7$ up to 50 qubits, as detailed in \changed{\ref{appendix-markov-sampling}}.

\changed{The precise number of samples required by SCU is given by Eq. \eqref{eq: SCU-number-samples}. The number of samples required by qDRIFT and the standard product formulas is also given by Eq. \eqref{eq: SCU-number-samples}, but with $\lambda = 1$, thus making the relative overhead of SCU $\lambda^2$. For example, in Fig. \ref{subfig:TFIM_CNOT_counts:qdrift}, the top of the blue CTS band corresponds to $\lambda=2$. Following Eq. \eqref{eq: SCU-number-samples}, this translates to requiring 4 times as many measurements as qDRIFT to achieve the same statistical precision. In \ref{appendix-TFIM-resource-analysis}, we provide a fixed-precision, total cost comparison between qDRIFT and CTS, accounting for both circuit depth and sampling costs.}

By properly including cross terms, our proposed algorithms model time evolution coherently, enabling empirical resource estimates with the spectral norm \cite{chen_concentration_2021}. In contrast, qDRIFT estimates require the diamond norm, which is challenging to compute for more than a few qubits. Following Campbell's paper \cite{campbell_random_2019}, we instead analytically bound the number of gates as $N \leq 2 \lambda_H^2 t^2 / \epsilon_{\diamond}$, where $\lambda_H$ is the $L_1$ norm of the Hamiltonian and $\epsilon_{\diamond}$ is the diamond norm error bound.  To enable comparisons in Fig. \ref{fig:TFIM_CNOT_counts}, we apply a bound on the diamond norm described in Refs.~\cite{chen_concentration_2021, kiss_importance_2023}. For a target unitary channel $\mathcal{U}(\rho)=U \rho U^{\dagger}$ and a probabilistic ensemble of unitary channels $\mathcal{V}_k(\rho)=V_k \rho V_k^{\dagger}$,
\begin{equation} \label{eq: diamond-spectral-norms-inequality}
\left\|\mathcal{U}-\sum_k p_k \mathcal{V}_k\right\|_{\diamond} \leq 2\left\|U-\sum_k p_k V_k\right\|.
\end{equation}
\changed{Accordingly, for each spectral target $\epsilon$ in Fig. \ref{fig:TFIM_CNOT_counts}, we evaluate qDRIFT with $\epsilon_{\diamond} = 2 \epsilon$ in Campbell's bound. We note that such analytic bounds likely overestimate the error, leading to higher resource estimates. We also note that uniformly weighted Hamiltonians like the TFIM are considered the worst case for qDRIFT \cite{campbell_random_2019}, so it may remain competitive for other systems.}

\section*{Acknowledgements}
This work is supported by an NSF CAREER Award under Grant No. NSF-ECCS-1944085 and the NSF CNS program under Grant No. 2247007. The authors acknowledge the use of IBM Quantum services for this work. The views expressed are those of the authors and do not reflect the official policy or position of IBM or the IBM Quantum team.

\section*{Author Contributions}
J.P. and S.E.S. developed the framework with guidance from P.N. J.P. designed the Hamiltonian simulation approaches with feedback from S.E.S. and P.N. J.P. implemented the GHZ experiment and computed the TFIM resource estimates with support from S.E.S. All authors contributed to the manuscript.




\bibliography{refs}

\clearpage
\onecolumngrid
\renewcommand{\thesubsection}{Supplementary Note~\arabic{subsection}}
\renewcommand{\theequation}{S\arabic{equation}}
\renewcommand{\thefigure}{S\arabic{figure}}
\renewcommand{\thetable}{S\arabic{table}}

\renewcommand{\theHequation}{S\arabic{equation}}
\renewcommand{\theHfigure}{S\arabic{figure}}
\renewcommand{\theHtable}{S\arabic{table}}
\renewcommand{\theHsubsection}{supp.\arabic{subsection}}

\setcounter{section}{0}
\setcounter{subsection}{0}
\setcounter{equation}{0}
\setcounter{figure}{0}
\setcounter{table}{0}
\section*{Supplementary Information}

\subsection{Generalized MQC for Damped GHZ} \label{appendix-damped-MQC}

In this section, we provide implementation details for the simulation of the damped GHZ state, building on the multiple quantum coherences (MQC) formalism used in previous works \cite{garttner_relating_2018, wei_verifying_2020, mooney_generation_2021}. A general density matrix can be decomposed in terms of its coherences $\hat{\rho}_{m}$ as $\rho = \sum_{m,q} \rho_{q, q-m}|q\rangle\langle q-m| = \sum_m \hat{\rho}_{m}$, and MQC provides a scalable technique to access these quantities experimentally. Here, we leverage MQC to compute the fidelity of a damped GHZ state $\rho$ relative to an ideal GHZ state $\rho^{\text{GHZ}}$. Eq. \eqref{eq: fidelity-decomp} in the main text expresses this fidelity as a sum over the populations $\hat{P} := \rho_{0^n, 0^n}+\rho_{1^n, 1^n}$ and the coherences $\hat{C} := \rho_{0^n, 1^n}+\rho_{1^n, 0^n}$. The populations are directly observable by simply preparing the damped GHZ state and measuring in the standard basis. We compute the coherences as $C = 2 \sqrt{I_n}$, where $I_m(\hat{\rho}) = \operatorname{Tr}[\hat{\rho}_m^\dagger \hat{\rho}_m]$ is an experimentally accessible metric called the multiple-quantum intensity.

As explained in \cite{garttner_relating_2018}, the derivation of MQC assumes unitary evolution. Gärttner et al. show that MQC also works for certain non-unitary channels, but amplitude damping as described here is not included due to its asymmetric decay to $\ket{0}$. In the following derivation, however, we show how the linearity of the SCU decomposition allows us to access the quantum coherences of the damped GHZ state through a generalized MQC scheme.

\begin{figure}[t!]
\centering
\includegraphics[width=0.53\columnwidth]{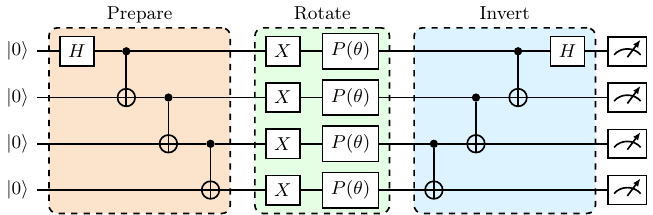} 
\caption{Multiple quantum coherences (MQC) circuit for 4 qubits, shown with no amplitude damping.}
\label{fig: MQC-circuit}
\end{figure}

Denote $\mathcal{E}_n$ as the channel which prepares the damped $n$-qubit GHZ state from $\ket{0^n}$, combining all CNOTs and single-qubit damping channels into a single operator. We decompose this channel as follows:
\begin{equation}
\begin{aligned}
    \mathcal{E}_n(\rho_0) & = \lambda\Biggl(\sum_j p_{jj} U_j^{} \rho_0 U_j^{\dagger} +\sum_{j, k<j} p_{j k}\left(e^{i \alpha_{j k}} U_j^{} \rho_0 U_k^{\dagger}+e^{-i \alpha_{j k}} U_k^{} \rho_0 U_j^{\dagger}\right)\Biggr) \\
    & = \lambda\Biggl(\sum_j p_{jj} \rho_j +\sum_{j, k<j} p_{j k}\left(e^{i \alpha_{j k}} \rho_{jk} + e^{-i \alpha_{j k}} \rho_{jk}^\dagger\right)\Biggr),
\end{aligned}
\end{equation}
where $\rho_j := U_j^{} \rho_0 U_j^{\dagger}$ and $\rho_{jk} := U_j^{} \rho_0 U_k^{\dagger}$. For the random-unitary terms, we use the standard MQC technique \cite{wei_verifying_2020} as depicted in Fig. \ref{fig: MQC-circuit}, while absorbing any stochastically sampled gates into $U_j$. After preparing $\rho_j$, we apply the rotation operator $R(\theta) = (P(\theta) X)^{\otimes n}$, invert the preparation via $U_j^\dagger$, and then measure the all-zero population $\hat{\rho_0}$. For the following derivation, observe that $R(\theta)\hat{\rho}_{j,m} R(\theta)^\dagger = e^{-im\theta} \hat{\rho}_{j,m}$. We thus measure the observable
\begin{equation}
\begin{aligned}
\expval{\hat{S}_{\theta,j}} & = \operatorname{Tr}\left[\rho_0 U_j^{\dagger} R(\theta) U_j^{} \rho_0 U_j^{\dagger} R(\theta)^\dagger U_j^{} \right]  \\
& = \operatorname{Tr}\left[(U_j^{} \rho_0 U_j^{\dagger}) R(\theta) (U_j^{} \rho_0 U_j^{\dagger}) R(\theta)^\dagger \right] \\
& = \operatorname{Tr}\left[\rho_j R(\theta) \rho_j R(\theta)^\dagger \right] \\
& = \sum_{m',m} \operatorname{Tr}\left[\rho_{j,m'} e^{-im\theta} \rho_{j,m}  \right] \\
& = \sum_{m} e^{-im\theta} \operatorname{Tr}\left[\rho_{j,-m} \rho_{j,m}  \right] \\
& = \sum_{m} e^{-im\theta} I_{m}(\rho_j).
\end{aligned}
\end{equation}

\pagebreak
For the cross terms, the derivation is similar. In order to properly invert the preparation, we exactly mirror the preparation gates by re-applying any sampled damping gates and using the same ancilla qubits. Overall, this yields
\begin{equation}
\begin{aligned}
\expval{\hat{S}_{\theta,jk}} & = \operatorname{Tr}\left[\rho_0 e^{-i \alpha_{j k}} U_j^{\dagger} R(\theta) e^{i \alpha_{j k}} U_j^{} \rho_0 U_k^{\dagger} R(\theta)^\dagger U_k^{} \right] + \operatorname{Tr}\left[\rho_0 e^{i \alpha_{j k}} U_k^{\dagger} R(\theta) e^{-i \alpha_{j k}} U_k^{} \rho_0 U_j^{\dagger} R(\theta)^\dagger U_j^{} \right]  \\
& = \operatorname{Tr}\left[(U_k^{} \rho_0 U_j^{\dagger}) R(\theta) (U_j^{} \rho_0 U_k^{\dagger}) R(\theta)^\dagger \right] + \operatorname{Tr}\left[(U_j^{} \rho_0 U_k^{\dagger}) R(\theta) (U_k^{} \rho_0 U_j^{\dagger}) R(\theta)^\dagger \right] \\
& = \operatorname{Tr}\left[\rho_{jk}^\dagger R(\theta) \rho_{jk} R(\theta)^\dagger \right] + \operatorname{Tr}\left[\rho_{jk} R(\theta) \rho_{jk}^\dagger R(\theta)^\dagger \right] \\
& = \sum_{m,m'} e^{-im\theta} \left(\operatorname{Tr}\left[\rho_{jk,m'}^\dagger \rho_{jk,m}  \right] + \operatorname{Tr}\left[\rho_{jk,m'} \rho_{jk,m}^\dagger  \right]\right) \\
& = \sum_{m} e^{-im\theta} I_{m}(\rho_{jk} + \rho_{jk}^\dagger) \\
& = \sum_{m} e^{-im\theta} I_{m}(e^{i \alpha_{j k}} \rho_{jk} + e^{-i \alpha_{j k}} \rho_{jk}^\dagger).
\end{aligned}
\end{equation}

Combining these two schemes in the framework of SCU, in total we measure the all-zero population signal $\hat{S}_\theta$ as
\begin{equation}
\begin{aligned}
    \expval{\hat{S}_\theta} & = \lambda\Biggl(\sum_j p_j \expval{\hat{S}_{\theta,j}} +\sum_{j, k<j} p_{j k} \expval{\hat{S}_{\theta,jk}} \Biggr) \\
    & = \lambda\Biggl(\sum_j p_j \sum_{m} e^{-im\theta} I_{m}(\rho_j) +\sum_{j, k<j} p_{j k} \sum_{m} e^{-im\theta} I_{m}(e^{i \alpha_{j k}} \rho_{jk} + e^{-i \alpha_{j k}} \rho_{jk}^\dagger) \Biggr) \\
    & = \sum_{m} e^{-im\theta}I_{m}\Biggl(\lambda \sum_j p_j \rho_j + \lambda \sum_{j, k<j} p_{j k} (e^{i \alpha_{j k}} \rho_{jk} + e^{-i \alpha_{j k}} \rho_{jk}^\dagger) \Biggr) \\
    & = \sum_{m} e^{-im\theta}I_{m} \left(\mathcal{E}_n(\rho_0)\right).
\end{aligned}
\end{equation}

Finally, by Fourier transforming the signals $\expval{S_\theta}$ shown in Fig. \ref{fig:MQC-signal-plot}, we obtain the multiple-quantum \changed{intensity
\begin{equation}
    I_{m} \left(\mathcal{E}_n(\rho_0)\right) = \frac{1}{N} \left|\sum_\theta e^{i m \theta} S_\theta\right|, 
\label{eq: fourier-MQC}
\end{equation} 
where $N$ is the number of measured angles.} The coherences are then $C = 2 \sqrt{I_n}$, which we use to compute the fidelity relative to the ideal GHZ state.

In the \changed{main article}, we compare the experimentally measured fidelities with those of an ideal damped GHZ state, i.e. one hypothetically prepared on a noise-free quantum device. For these values, we use the following analytic calculations. The $n$-qubit GHZ state preparation involves $n-1$ instances of CNOT-induced amplitude damping. As a result, the all-one population is reduced by a factor of $(1-p)^{n-1}$, and the coherences are reduced by $(1-p)^{(n-1)/2}$. Using these damped values in Eq. \eqref{eq: fidelity-decomp} in the main text gives an analytically predicted fidelity of $F = \frac{1}{4}\left(1+(1-p)^{(n-1) / 2}\right)^2$.

\subsection{Damped GHZ Resource Analysis} \label{appendix-damping-resources}
\begin{figure}[h!]
\centering
\includegraphics[width=0.42\columnwidth]{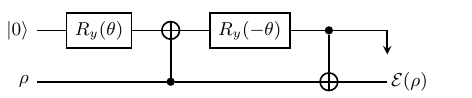} 
\caption{Direct implementation of amplitude damping channel $\mathcal{E}$ with strength parameter $p = \cos{(\theta)}$.}
\label{fig:damping-circuit}
\end{figure}

In this section, we compare the resource costs of our SCU-based approach to the direct circuit implementation shown in Fig. \ref{fig:damping-circuit}. Compared to previously proposed circuits for amplitude damping \cite{rost_simulation_2020}, we designed this circuit to be CNOT-efficient to make a fairer comparison with our stochastic approach. In Table \ref{table:damping-resources}, we show that SCU requires a sampling overhead of $(1+p)^2$ per instance but has a reduced average CNOT cost of $\frac{2p}{1+p}$.  SCU thus offers significant reductions in circuit depth, especially pronounced for weak damping parameters $p$. As an example, consider our eight-qubit GHZ simulation with $p=5\%$. For this case, the SCU approach requires a sampling overhead of $\lambda^2 =  1.98$.  However, including damping in the eight-qubit MQC circuit uses 28 CNOT gates with the direct approach compared to an average of 1.27 CNOT gates with SCU. Thus, with minimal sampling overhead, our SCU-based approach reduces these gate costs by over an order of magnitude.

For strong damping parameters $p$, the average CNOT cost with SCU is at most 1, still an improvement over the direct circuit. However, as the sampling overhead per instance increases, the scalability of this method suffers and eventually becomes impractical. In general, this trade-off between depth reductions and sampling overhead \changed{should be balanced for particular applications.}

\changed{The damped GHZ preparation via SCU represents the simplest dilation, as it involves only a single qubit. If we were to treat this with a traditional LCU approach, the results would be much worse. In Table \ref{table:damping-resources}, we see that an LCU dilation requires a smaller sampling overhead of $(1+p)$, but its depth grows with the number of Kraus operators. Specifically, each damping instance requires 2 ancilla qubits implementing variants of Toffoli operations for 4 unitaries, compiling to at least 24 CNOTs before accounting for the ancilla state preparation. While this circuit depth is independent of $p$, the success probability of LCU still influences the requisite shot overhead.}

\renewcommand{\arraystretch}{1.5}
\begin{table}[h]
    \centering
    \begin{tabular}{c|c|c|c}
    \textbf{Method} & \textbf{$\expval{N_{CX}}$} & \textbf{Overhead} & \textbf{Ancilla} \\ \hline
    Direct & $2$ & $1$ & $1$ \\
    Stochastic & $\frac{2p}{1+p} \leq 1$ & $(1+p)^2$ & $1$ \\
    LCU & 24 & $1+p$ & 2
    \end{tabular}
    \caption{\changed{Comparison of resource counts for different means of applying the damped GHZ operations.}}
    \label{table:damping-resources}
\end{table}

\subsection{Markov Chain Sampling} \label{appendix-markov-sampling}
In order for our SCU-based algorithms to be useful in early demonstrations of quantum advantage, we need to classically compute stochastic correction operators for reasonably large system sizes $n$, i.e. approaching 100 qubits. With this goal in mind, here we limit our truncation order to at most $M = 4$; this is merely a heuristic choice with which we can quickly compute corrections up to $n \approx 50$ qubits on our 2023 personal computers. However, by borrowing ideas from Markov chains, we can use even higher-order operators without the need to fully expand and simplify them. 

As an example, product formula remainder expressions can contain operators of the form $A = \sum_i^{O(n)} c_i P_i$ and $B = \sum_j^{O(n)} c_j P_j$. For reasonably large $n$, we can use Pauli algebra to fully compute second-order terms like $AB$ or even fourth-order terms like $A^2 B A$. In contrast, seventh-order terms such as $A^4 B^3$ quickly become too costly to expand in the same way. We propose partitioning such terms into multiple layers of sampling, which in aggregate approximate the full operator. For this case, we 1) separately expand $A^4 = \sum_l c_l P_l$ and $B^3 = \sum_m c_m P_m$, 2) sample each as independent ensembles $\{P_l\}$ and $\{P_m\}$, and 3) approximate $A^4 B^3$ as the product of sampled operators $\{P_l P_m\}$. This yields an unbiased estimator but generally increases the $L_1$ norm compared to full expansion.

For the enhanced second-order product formula, this structure-agnostic approach allows us to achieve a spectral error corresponding to truncation order $M=7$ while still managing to empirically compute the overhead bounds up to 50 qubits. The downside is that we miss out on potential simplifications between the various remainder terms, ultimately increasing the sampling overhead. For example, applying this scheme to all remainders of the first-order product formula would effectively ignore the commutator scaling of the second-order correction, causing the normalization constant $\lambda$ to scale worse with respect to system size.

\subsection{Transverse Field Ising Model: Resource Analysis} \label{appendix-TFIM-resource-analysis}
After computing the number of time steps $r$, we convert to CNOT cost. For the standard product formulas, we simply multiply the number of two-qubit Pauli exponentials by a factor of two. For qDRIFT, we multiply the total number of gates by the probability of sampling a two-qubit Pauli exponential, again rescaling this by a factor of two. For the enhanced product formulas, the dominantly sampled operators are the standard product formulas $\mathcal{S}_p$. For a time step $t/r$, the probability of sampling $\mathcal{S}_p$ is $p_0 = 1/\mu(t) = \lambda^{-1/2r}$, and the probability of sampling first $\mathcal{S}_p$ for the left operator and then $\mathcal{S}_p^\dagger$ for the right operator is $p_0^2 = \lambda^{-1/r}$. The probability of any other event is thus $p_1 = 1 - \lambda^{-1/r}$. Over all time steps $r$, the expected number of sampled terms $\expval{N_1}$ other than the standard product formula $\mathcal{S}_p \rho \mathcal{S}_p^\dagger$ is simply
\begin{equation}
    \expval{N_1} = r p_1 = r (1-\lambda^{-1/r}).
\end{equation}
Asymptotically, for large r, we obtain
\begin{equation}
    \expval{N_1} \sim \ln{(\lambda)}.
\end{equation}
We conclude that the gate cost per time step is negligibly increased compared to the standard product formula approach.

\begin{figure}[h!]
\centering
\includegraphics[width=0.35\columnwidth]{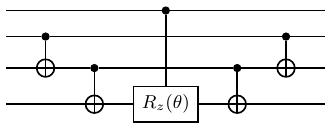} 
\caption{Circuit for $C(e^{-i \frac{\theta}{2} (Z^{\otimes n})})$, shown here for $n=3$. This generalizes to any Pauli string rotation via local similarity transformations.}
\label{fig:CZZZ}
\end{figure}

For CTS, the dominantly sampled gates will be Pauli exponentials from $I - iHt$, and the sampled left and right gates will typically differ. For the TFIM with $J=h=1$, approximately half of the sampled gates will be $e^{-i \theta Z Z}$  and the other half will be  $e^{-i \theta X}$. A controlled single-qubit Pauli exponential is locally equivalent to $C(R_z(\theta))$, which compiles to two CNOT gates. As shown in Fig. \ref{fig:CZZZ}, a controlled two-qubit Pauli exponential can be implemented with four CNOT gates. Thus on average, CTS uses approximately six CNOT gates per time step.

To minimize the need for ancilla qubits, one can use mid-circuit measurements, repeatedly resetting and reusing the same physical qubits. In principle, this approach requires only a single ancilla qubit for the entire simulation. Depending on the device capabilities, however, such measurements could create a bottleneck; a practical solution is to use a pool of ancilla qubits, thus lowering the frequency of measurements per individual qubit.

\changed{\subsubsection{Total cost comparison}}

\changed{We now provide a fixed-precision, total cost comparison between qDRIFT and CTS for simulating the $50$-qubit TFIM, accounting for both circuit depth and sampling costs. The results are shown in Table \ref{tab:fixed-epsilon-qDRIFT-CTS-costs}. We allocate a fixed total error budget of
\begin{equation} \label{eq: epsilon-total}
    \epsilon_{tot} = \epsilon_{stat} + \epsilon,
\end{equation}
where $\epsilon$ is the spectral error and $\epsilon_{stat}$ is the statistical sampling precision, as seen in Eq. \eqref{eq: SCU-number-samples} in the main text for SCU-based methods like CTS. The same equation describes the sampling costs of qDRIFT but with $\lambda=1$. Explicitly, the number of samples required by qDRIFT is 
\begin{equation} \label{eq: qDRIFT-number-samples}
N_{\mathrm{qDRIFT}} \;\ge\; \frac{2\,\|O\|^{2}}{\epsilon_{stat}^{2}}\,
\ln\!\frac{2}{\delta},
\end{equation}
where $\|O\|$ denotes the spectral norm of the observable and $\delta$ denotes the failure probability.}

\begin{table}[t!]
\centering
\begin{tabular}{c|c|c|c|c}
\hline
\textbf{$\varepsilon_{\mathrm{tot}}$} & \textbf{Method} & \textbf{\# Samples} & \textbf{CNOTs per Circuit} & \textbf{Total CNOT Cost} \\
\hline
\multirow{2}{*}{$10^{-3}$} & qDRIFT & \num{1.660e7} & \num{7.277e10} & \num{1.208e18} \\
\hhline{~----}
 & CTS & \num{2.951e7} & \num{4.109e8} & \num{1.213e16} \\
\hline
\multirow{2}{*}{$10^{-6}$} & qDRIFT & \num{1.660e13} & \num{7.277e13} & \num{1.208e27} \\
\hhline{~----}
 & CTS & \num{2.951e13} & \num{4.109e8} & \num{1.213e22} \\
\hline
\end{tabular}
\caption{\changed{Resource cost comparison between qDRIFT and CTS for simulating the $50$-qubit TFIM with evolution time $t = 50$ and a total error budget of $\epsilon_{tot}$. We consider observables of unit norm $\|O\| = 1$ (such as Pauli observables) and fix the failure probability at $\delta = 0.05$ and the normalization constant at $\lambda = 2$. The total CNOT cost is the product of the number of samples and the CNOTs per circuit.}}
\label{tab:fixed-epsilon-qDRIFT-CTS-costs}
\end{table}

\changed{For qDRIFT, we determine $\epsilon_{stat}$ and $\epsilon$ by optimizing the product of the number of time steps \cite{campbell_random_2019} and the number of samples:
\begin{equation}
    \left(\frac{\lambda_H^2 t^2}{\epsilon}\right) \left(\frac{2\,\|O\|^{2}}{\epsilon_{stat}^{2}}\,
\ln\!\frac{2}{\delta} \right).
\end{equation}
As in the main text, here we set $\epsilon_{\diamond} = 2 \epsilon$ following the norm inequality in Eq. \eqref{eq: diamond-spectral-norms-inequality}. By substituting $\epsilon = \epsilon_{tot} - \epsilon_{stat}$, we can minimize the resulting expression by setting the first derivative with respect to $\epsilon_{stat}$ to $0$. We find the optimal sampling allocation for qDRIFT to be
\begin{equation}
    \epsilon = \frac{\epsilon_{tot}}{3} \;\;\text{and} \;\; \epsilon_{stat} = \frac{2 \,\epsilon_{tot}}{3}.
\end{equation}
Thus, in Table \ref{tab:fixed-epsilon-qDRIFT-CTS-costs}, the number of samples is given by Eq. \eqref{eq: qDRIFT-number-samples} with $\epsilon_{stat} = \frac{2 \,\epsilon_{tot}}{3}$, and the number of time steps is $3 \lambda_H^2 t^2 / \epsilon_{tot}$. For the TFIM, we require two CNOTs every time we sample the Pauli string $ZZ$, which occurs with probability $J(n-1) / (hn + J(n-1))$. Thus, the average CNOTs per circuit for qDRIFT in Table \ref{tab:fixed-epsilon-qDRIFT-CTS-costs} is
\begin{equation}
    \left(\frac{2J(n-1)}{hn + J(n-1)}\right) \left(\frac{3 \lambda_H^2 t^2}{\epsilon_{tot}} \right),
\end{equation}
where $n$ is the number of qubits.}

\changed{For CTS, we fix the normalization constant to be $\lambda=2$ in Table \ref{tab:fixed-epsilon-qDRIFT-CTS-costs}. Because the sampling precision is the dominant constraint, the spectral error $\epsilon$ is negligibly small compared to the total error budgets considered of $\epsilon_{tot} \in \{10^{-3},10^{-6}\}$. This can be seen in Fig. \ref{subfig:TFIM_CNOT_counts:qdrift}, where the resource requirements of $\lambda=2$ greatly exceed those of the spectral error constraints. Thus, for CTS, we allocate the full error budget to $\epsilon_{stat}$, and the CNOTs per circuit in Table \ref{tab:fixed-epsilon-qDRIFT-CTS-costs} directly match the values in Fig. \ref{subfig:TFIM_CNOT_counts:qdrift}. The number of samples are computed from Eq. \eqref{eq: SCU-number-samples} in the main text. The techniques in this illustrative example can be applied to the stochastically enhanced product formulas as well, which demonstrate even better performance in Fig. \ref{fig:TFIM_CNOT_counts}.}
\vspace{3mm}

\subsection{Proof of Theorem 1} \label{subsection: appendix-theorem1-proof}
First, we Taylor expand $e^{-iHt}$ and separate the terms into real and imaginary subsets.
\begin{equation}
\begin{aligned}
& e^{-iHt} = \sum_{l=0}^{\infty} \frac{1}{l!}(-itH)^l \\
& = \sum_{m=1}^{\infty} \frac{(-1)^m}{(2m)!} (tH)^{2m} + I + i \sum_{m'=0}^{\infty} \frac{(-1)^{m'+1}}{(2m'+1)!} (tH)^{2m'+1}.
\end{aligned}
\end{equation}
Now substitute the following Pauli decompositions of each sum:
\begin{equation} \label{eq:even_odd_sums}
\begin{aligned}
\sum_{m=1}^{\infty} \frac{(-1)^m}{(2m)!} (tH)^{2m} & = \sum_j c_j P_j = L_c \sum_j p_j P'_j \\
\sum_{m'=0}^{\infty} \frac{(-1)^{m'+1}}{(2m'+1)!} (tH)^{2m'+1} & = \sum_k c_k P_k = L_s \sum_k p_k P'_k.
\end{aligned}
\end{equation}
Because all powers of $H$ are Hermitian, the coefficients $c_i$ are strictly real. We absorb any negative signs into the operators $P'_i := \textrm{sign}(c_i) P_i$, and normalize by the $L_1$ norms $L_c := \sum_j |c_j|$ and $L_s := \sum_k |c_k|$. This ensures that $\sum_i p_i = 1$, thus creating convex sums of unitaries. Continuing, we get:
\begin{equation}
\begin{aligned}
e^{-iHt} & = L_c \sum_j p_j P'_j + I + i L_s \sum_k p_k P'_k \\
& = L_c \sum_j p_j P'_j + \sum_k p_k (I + i L_s P'_k) \\
& = L_c \sum_j p_j P'_j + \sqrt{1+L_s^2} \sum_k p_k e^{i \theta P'_k}.
\end{aligned}
\end{equation}
In the final step, we use the substitution $I + i L_s P'_k = \sqrt{1+L_s^2} e^{i \theta P'_k}$, where $\theta := \arcsec{(\sqrt{1+L_s^2})}$. This step was inspired by the excellent work of Wan et al. \cite[Appendix C]{wan_randomized_2022}. Finally, to prove the upper bounds on the $L_1$ norms, notice that $\sum_{m=1}^{\infty} \frac{1}{(2m)!} (tH)^{2m} = \cosh{(tH)} - 1$ and $\sum_{m'=0}^{\infty} \frac{1}{(2m'+1)!} (tH)^{2m'+1} = \sinh{(tH)}$. Because $|H^n|_1 \leq |H|_1^n$, we conclude that $L_c \leq \cosh{(t|H|_1)} - 1$ and $L_s \leq \sinh{(t|H|_1)}$. Importantly, these bounds establish that the leading-order scaling of the total $L_1$ norm is quadratic in time, i.e. $L_c + \sqrt{1+L_s^2} \sim 1 + t^2|H|_1^2$.

\subsection{Worst-Case Complexity Analysis} \label{subsection:worst-case-complexity}
Consider a general quantum channel $\mathcal{E}$, with the following Kraus decomposition: 
\begin{equation}
\mathcal{E}(\rho)=\sum_{i}^\kappa K_i \rho K_i^{\dagger}.
\end{equation}
Any Kraus operator $K_i$ can be written as a linear combination of unitary matrices, $K_i = \sum_{j}^{M_i} c_{ij} U_{ij}$. For instance, the Pauli strings form an orthonormal basis of operators. Then, we obtain:
\begin{equation} \label{kraus_unitaries}
\begin{aligned}
    \mathcal{E}(\rho) & =\sum_{ijk} c_{ij} c_{ik}^* U_{ij}^{} \rho U_{ik}^\dagger. \\
\end{aligned}
\end{equation}

\noindent \textbf{Theorem 2}: Suppose that each $\kappa$ Kraus operator $K_i$ decomposes into $M_i$ unitaries $U_{ij}$, and that $\sum_i^\kappa M_i = N$. Then, the $L_1$ sampling norm of the coefficients in \changed{Eq.} \eqref{kraus_unitaries} is at most $M^* := \textrm{max}(\{M_i\})$. \\

\noindent \textbf{Proof}: First, we directly compute the $L_1$ norm of the coefficients in \changed{Eq.} \eqref{kraus_unitaries}:
\begin{equation}
\begin{aligned}
|\mathcal{E}|_1 & = \sum_{ijk} |c_{ij} c_{ik}^*| \\
& = \sum_i \left(\sum_{j}^{M_i} |c_{ij}| \sum_{k}^{M_i} |c_{ik}|\right) \\
& = \sum_i \left(\sum_j^{M_i} |c_{ij}|\right)^2.
\end{aligned}
\end{equation}
Now, use the property that $\mathcal{E}$ is trace-preserving, i.e. $\sum_i K_i^\dagger K_i = I$:
\begin{equation}
\begin{aligned}
I & = \sum_{ijk} c_{ij}^* c_{ik} U_{ij}^\dagger U_{ik}^{} \\
& = \sum_{ij} |c_{ij}|^2 I + \sum_{ij,k \neq j} c_{ij}^* c_{ik} U_{ij}^\dagger U_{ik}^{}.
\end{aligned}
\end{equation}
Thus, the second sum must be $0$, but more importantly,
\begin{equation}
\sum_{ij} |c_{ij}|^2 = 1.
\end{equation}
In other words, the $L_2$ norm of the set $\{|c_{ij}|\}$ is $1$. According to the Cauchy-Schwarz inequality, for complex numbers $u_i$ and $v_i$,
\begin{equation}
    \left|\sum_{i=1}^n u_i v_i^* \right|^2 \leq \sum_{j=1}^n |u_j|^2 \sum_{k=1}^n |v_k|^2.
\end{equation}
Setting $v_i = 1 \; \forall i$, we obtain the following identity:
\begin{equation}
    \left|\sum_{i=1}^n u_i \right|^2 \leq n \sum_{j=1}^n |u_j|^2.
\end{equation}
This is exactly what we need -- by knowing the $L_2$ norm, we can set an upper limit on the $L_1$ norm. For our application, we set $u_i = |c_{ij}|$ to obtain the following inequality, valid for all $i$:
\begin{equation}
    \left(\sum_{j}^{M_i} |c_{ij}|\right)^2 \leq M_i \sum_{j}^{M_i} |c_{ij}|^2.
\end{equation}
Thus,
\begin{equation}
    |\mathcal{E}|_1 \leq \sum_i M_i \sum_{j}^{M_i} |c_{ij}|^2 \leq M^* \sum_{ij} |c_{ij}|^2.
\end{equation}
Because the $L_2$ norm is 1, we get our final result:
\begin{equation}
    |\mathcal{E}|_1 \leq M^*.
\end{equation}

\end{document}